\documentclass[12pt,preprint]{aastex}
                                                                                                   
\newcommand{\msun}{$M_{\sun}$}

\slugcomment{To appear in ApJ}
                                                                                                   
\shorttitle{SN-driven HI distribution}
\shortauthors{Hodge \& Deshpande}
                                                                                                   
\begin{document}
                                                                                                   
                                                                                                   
\title{HI Density Distribution Driven by Supernovae:\\
A Simulation Study}


\author{Jacqueline A. Hodge\altaffilmark{1}}
\affil{California Polytechnic State University, San Luis Obispo CA 93407, USA}
\email{jhodge@calpoly.edu}

\and
                                                                                                   
\author{Avinash~A.~Deshpande\altaffilmark{2}}
\affil{NAIC/Arecibo Observatory, HC3 Box 53995, Arecibo, Puerto Rico 00612}
\email{desh@naic.edu}
                                                                                                   
                                                                                                   
\altaffiltext{1}{REU program, Arecibo Observatory, HC3 Box 53995, Arecibo, Puerto Rico 00612\\
Present address: Department of Physics, University of California, One Shields Avenue, Davis, CA 95616; email: hodge@physics.ucdavis.edu}
\altaffiltext{2}{Raman Research Institute, Bangalore 560 080 INDIA; e-mail: desh@rri.res.in}

                                                                                                   
\begin{abstract}
We model the complex distribution of atomic hydrogen (HI) in the 
interstellar medium (ISM) assuming that it is driven entirely by 
supernovae (SN). We develop and assess two different
models. In the first approach, the simulated volume is randomly 
populated with non-overlapping voids of a range of sizes. This 
may relate to a snapshot distribution of supernova-remnant voids, 
although somewhat artificially constrained by the non-overlap 
criterion. In the second approach, a simplified time evolution 
(considering momentum conservation as the only governing constraint 
during interactions) is followed as SN populate the space with the 
associated input mass and energy. 

We describe these simulations and present our results in 
the form of images of the mass and velocity distributions and
the associated power spectra. The latter are compared with trends
indicated by available observations. In both approaches, we find 
remarkable correspondence with the observed statistical description 
of well-studied components of the ISM, wherein the spatial spectra
have been found to show significant deviations from the Kolmogorov 
spectrum. One of the key indications from this study, regardless of 
whether or not the SN-induced turbulence is the dominant process in
the ISM, is that the apparent non-Kolmogorov spectral characteristics 
(of HI and/or electron column density across thick or thin screens) 
needed to explain related observations may not at all be in conflict 
with the underlying turbulence (i.e. the velocity structure) 
being of Kolmogorov nature. We briefly discuss the limitations of
our simulations and the various implications of our results.
\end{abstract}

                                                                                                   
                                                                                                   
\keywords{Galaxies: ISM --- ISM: Kinematics and Dynamics,
Turbulence, HI, Structure --- Radio Lines: ISM --- 
Supernovae: Explosion --- Scattering --- Pulsars: General}

\noindent{\section{Introduction}}

  The non-uniform and turbulent nature of the interstellar
medium (ISM) has long been evident from scattering/scintillation studies in the
direction of pulsars and compact extra-galactic sources, as well as from HI
emission/absorption measurements. Power spectra of the observed 
column density distribution of neutral hydrogen (HI) in the 
ISM reveal its fractal nature.  
Some estimates of the slope of spatial power
spectra for the absorbing HI distribution in the Galaxy show a range from --3.0 
(Crovisier and Dickey 1983) to
--2.75 $\pm$ 0.25 (Deshpande, Dwarakanath \& Goss 2001), while Green (1993)
found a range from --2.2 to --3.0 for the slopes from HI emission spectra,
indicating steepening with distance. 
Several recent statistical studies of galactic and extra-galactic HI on large scales
(e.g. Stanimirovic et al. 1999, Dickey et al. 2001, Elmegreen et al. 2001, 
Stanimirovic \& Lazarian 2001) also reveal power-law slopes characterizing
the spatial power spectra that are consistent with 
the above-mentioned range of values. The prediction of Lazarian \& Pogosyan (2000), 
that the spatial spectra should steepen with the increasing thickness of the velocity
slice, was confirmed by some of these analyses (e.g. Stanimirovic \& Lazarian 2001).
The question we would like to address here is why the structure of the HI is 
like this in a statistical sense.  

A prominent source of energy input driving the ISM evolution is
thought to be supernovae (SN). Many researchers have attempted
detailed MHD simulations of a supernova-driven ISM to explore this aspect.
For example, Korpi et al. (1999) used a three-dimensional, non-ideal MHD
model and monitored several physical parameters.  
There have been several other detailed simulation studies exploring 
MHD turbulence in different regimes (Cho, Lazarian \& Vishniac 2002; 
Wada et al. 2002; Vollmer \& Beckert 2002; Cho \& Lazarian 2003; 
Miville-Deschenes, Levrier \& Falgarone 2003; Slyz et al. 2005; and
Mac Low et al. 2005).

Our simulation study takes
a highly simplified view of the ISM and monitors only the statistical
description of the density and velocity distributions.  As a first step
toward assessing what the essential features of the processes underlying
a supernova-driven ISM are, we retain the kinematical and structural
components alone and assume that these are entirely determined by the
expanding shells of SNRs.  

We explore two basic approaches in our modeling of a
SN-driven ISM; I) a simple simulation of an equivalent snapshot 
distribution, and 
II) a more complex simulation incorporating time evolution.  Sections 2 and 3
describe the details and limitations of these two modeling approaches separately,
while Section 4 is dedicated to the analysis of both of the models.  Results are
discussed in Section 5, and our conclusions on the implications of the two
models are presented in Section 6.

\noindent{\section{Snapshot Distribution}}

With the goal of keeping our model as simple as possible, we first model
resultant contributions from supernovae
directly as an ensemble of simple, non-overlapping, spherical bubbles 
(or voids) in what is  otherwise a homogeneous ISM.
We use a three-dimensional matrix to simulate a region in space that we
wish to populate with a large number of bubbles of different radii. 
In this picture, the bubbles of different sizes
relate to supernova remnants (SNRs) of correspondingly different ages 
(and in different
phases of their evolution) that have resulted from supernovae at different epochs.
This method effectively generates a snapshot
in time, and we invoke the ergodicity theorem to justify the basis of
such a simulation.  Thus, while the bubbles do not evolve and grow iteratively
in this approach,
we attempt to achieve the same end result by simply generating simultaneously
a random distribution of locations and sizes.  For this first simulation,
each of the volume cells carry a binary tag (i.e. 0 or 1) indicating whether the given cell 
(or pixel) is inside a bubble or outside in the undisturbed ISM.  We wrap 
the respective sides (\& corners) of the array, ensuring continuation of the bubbles across 
them, to avoid any undesirable edge-effects, e.g. discontinuities and consequent
``mass loss" at the edges. This procedure is indeed consistent with the wrapping that
is implicit in the Fourier analysis of this data, a step that is performed subsequently.
We use a random number generator (with uniform distribution) to determine the coordinates 
and radius of each
bubble.  Several parameters, such as the minimum bubble radius, the
maximum bubble radius, and the total number of bubbles, are retained as
user selectable variables.  Checks are made, and iterations are skipped if needed
to ensure that new bubbles do not overlap with any of the previously input bubbles.  
We note that this non-overlap constraint is unphysical, but was unavoidable
in order to ensure that the simulated volume was not entirely emptied at any stage.
Naturally, the distribution of radii for included bubbles would deviate significantly
from a uniform distribution if not explicitly constrained, 
such that the number of bubbles dropped rapidly with increasing radius. 
Such a decline in the number of large holes may in fact 
be expected if the sources of energy input are strongly clustered, and
are of a wide range of energies.
An image resulting from this simulation is shown
in Figure~\ref{fig:bubble_image}.
Similar simulations are repeated a) by constraining the resultant distribution of the 
radii to be uniform, and/or b) with smooth-edged bubbles (with density falling 
towards center as $(1-cos(\pi r/r_0))/2$, where $r_0$ is the bubble radius, 
and $r$ is the distance from the bubble center).

\noindent{\section{Time Evolution}

The second approach to the simulation process involves incorporating a time axis
and allowing planted supernovae and the consequent SNRs to evolve.  
We make several assumptions for this simple model.  
To begin with, we include only Type II supernovae in our simulation.
We use an average galactic rate of 1/44 yr$^{-1}$ 
(Tammann, Loeffler \& Schroeder 1994) and
consider a uniform distribution across a 20\% range about this rate,
although Type II supernovae are correlated in both space and time.
The progenitor stars are assumed to have
masses in the range 8-20 solar masses (\msun), and their non-uniform
distribution within this range is modeled in consistency with 
the initial mass function as
given by Wheeler, Miller \& Scalo (1980) and approximated in our range of interest
by a power law.  As for the compact remnant mass, we assume a value of 1.4 \msun.
We then consider two types of ejecta, diffuse and clumpy (Willingale et al.
2003), with different velocities, 
and use the relevant values estimated by Willingale et al based on the Cas-A data.  
Finally, we assume that the rest of the mass
(i.e., M[progenitor] -- M[compact-remnant] -- M[ejecta]) 
was lost prior to the explosion by the star's stellar wind and other mass loss
mechanisms. We will denote this component by M[preSN-loss].

For this simulation, we use four 3-dimensional arrays to store
the spatial distribution of 
the mass and the 3-d velocity that evolve as a function of time.
The spatial resolution (the volume corresponding to a pixel or a cell) and 
the dynamic range of scales  
are chosen by specifying, in turn, suitable values for both the
spatial extent represented by the entire cube and the number of pixels in
the desired storage arrays. 
We initialize the mass-cube with a starting ISM density, assumed
uniform and generally of the order of 0.6 atoms/cm$^3$.  This value is
several times smaller than the average value that the ISM density would 
attain after subsequent mass input from supernovae. 

To initiate a supernova within the simulated
space, we choose a random pixel location in 3-d, and input the mass \& velocity distribution
associated with the supernova in the form of an expanding spherical 
shell.
The mass in this shell consists of the ejecta mass, M[ejecta],
as well as the mass lost during the pre-supernova phase, M[preSN-loss].
The inclusion
of this latter component is justified if the distance traveled by 
this material, which is expelled in the pre-SN
phase, is within the above mentioned cell size. This is indeed so given the typical cell
sizes in our simulations, and in fact we adjust the spatial resolution to ensure
this to be the case. 
The magnitude of the true initial velocity is estimated by conserving 
total kinetic energy 
between the two different types of ejecta (with assumed masses and velocities, 
see Willingale et al. 2003 for details), and the mass lost during the 
pre-supernova phase (i.e., M[preSN-loss]).
We use a basic template cube (3 x 3 x 3 pixels) to
store the 3-d velocity vectors directed radially outward and initially
normalized to a unit length.  We then multiply these vectors by the magnitude of the
true initial velocity as computed above.
To calculate the initial mass distribution in the shell, we distribute the
available mass into the template cube proportionally to the surface area of a
sphere centered in the middle pixel and of radius 1 pixel.  
As can be imagined, such a cube of cells is too coarse to capture the desired
spherical symmetry of the shell. As discussed later, we have checked the sensitivity
of our results to the size of this cube (i.e. the shell radius), and find that
the final results are practically unaltered even when we used a 5 x 5 x 5 matrix
as a template cube to model the seed shell.

After the
first explosion is initiated, we evolve the simulation through a large
number of iterations, each with a dynamic time step.  To calculate what
the time step should be for a particular iteration, we search the velocity
arrays for the maximum magnitude and use this to
set the time step such that the ``movement" at this speed is restricted
to less than one pixel.  This procedure
ensures that, for a given iteration, the range of influence of any 
pixel will never extend beyond a 3 x 3 x 3 cube around that pixel.  Thus, for each
iteration, we scan through the pixel set, ``evolve" the contribution of each
pixel by considering its interaction with matter within
a small 3 x 3 x 3 influence cube around it, and store/add that resultant contribution
to a buffer set of arrays for mass and 3-d velocity.

Within this smaller cube of {\it influence}, we calculate the movement of the mass 
and use a spread function to decrease the effects of the quantization
introduced by the finite pixel size.  Let $\hat{d}$ be a unit vector 
defining the direction of the destination pixel w.r.t. the central (or source) pixel, and
let D be the associated distance. The displacement $\delta$D in that direction is 
then simply $(\vec{V}.\hat{d}) \delta t$, where $\vec{V}$ is the velocity vector
and $\delta t$ is the time step. We restrict the spread of the central
pixel's contribution to destination pixels with positive values of $\delta$D,
i.e. to only those pixels that can be reached in the `forward' direction.
The  fractional displacement f = $\delta$D/D  determines how the original mass
is to be shared between the original pixel and the destination pixel.
Thus the fraction of the source mass transported to a destination pixel $i$ 
is $W_i.f$, where
$W_i=A_i/(\sum_i A_i)$,  and $A_i = (\vec{V}.\hat{d}/|V|)$ if the dot-product is
positive, else $A_i = 0$.
This consideration allows us to weight the mass
transfer in the forward direction while still spreading it adequately laterally.
Total time is kept track of throughout the iterations, and another
explosion is initiated whenever the time after the last explosion reaches
a pre-computed interval. This entire process is
carried out until the input ``run time" is complete, at which point the
program switches into the ``evolution-only" phase.  In this phase, we allow the
iterations to continue for a certain defined length of time, but
disallow any new supernova.  This phase of the simulation ensures the avoidance of
any extreme densities (both high \& low) due to a more recent SN/SNR (closer to the 
end of the first phase) that has not yet had time to evolve and 
interact with the surrounding matter. As our simulation is intended to 
study the structure of
the ISM, not the stars within it, this provision is necessary to make sure
recent explosions do not skew our results. 
Figure~\ref{fig:col_den_sne_image} shows a sample result from a full 
simulation in the form of a 2-d distribution of column density.

\noindent{\section{Analysis \& results}}

  Throughout our simulation runs (numbering ten or more in each of the cases
explored), particularly in the ``Time Evolution" approach, 
we closely monitor a variety of parameters and distributions. Typically,
for the evolutionary model, every 50 or so
iterations, we output the results in the form of mass (or density) \& velocity distributions
and
obtain power-spectral descriptions for these. 
For estimating the power spectra, in general, a given 2-d distribution is first 
stripped of
its ``mean", and  then Fourier transformed. The resultant 2-d power spectrum is
azimuthally averaged to obtain a 1-d description of the power as a function of
the spatial frequency (extending up to the sides, and not the corners).
The power estimates at
higher spatial frequencies benefit increasingly from the better statistics
implicit in azimuthal averaging as a function of radius.
These spectra are displayed on a log-log scale and best-fit power-law slopes
(spectral indices) are computed over 3 (partly overlapping) ranges of the
spatial frequencies to allow for the possibility that a given spectrum
may not be adequately described by a single power-law.

In the ``Snapshot" simulation, we mainly monitor the column density 
distribution.  We also examine the density distributions (in, say, the 
X-Y plane) corresponding to thin slices at various depths (i.e. at various Z 
values).  Here, the power spectra from several slices (typically 32) were 
combined to obtain a better average description.  

For the ``Time Evolution" simulation, we extend our analysis to include 
two more data sets.  The Z-velocity distribution is monitored (for thin Z-
slices), as well as the column density distribution as a function of 
velocity.  The latter can be most directly compared to observational results. 

\noindent{\subsection{The Snapshot Distribution}}

The analysis of the column density distribution in Figure~\ref{fig:bubble_image} 
was performed following the
above procedure, and the resultant power spectrum is shown in 
Figure~\ref{fig:bubble_spectrum}a.
As noted earlier, the underlying 3-d distribution includes a large number 
(500,000) of bubbles/voids. 
The spatial scale or the size of the simulated box is not explicitly defined, however
an approximate scale of 1 kpc may be associated with the width of the simulated box.
The best-fit power-law slope for the linear
portion of this log-log plot is close to -2.9, steeper than $-8/3$. 
Interestingly, the spectral steepness decreased monotonically (starting from a slope 
of about $-4$ for a single bubble) as the number of bubbles increased. This trend
is understood as due to a) the improved statistics, in general,
and b) the increased weightage for finer structure, in particular.
Also, importantly, the rate of change of the spectral slope is 
observed to decrease steadily.
The very slow decrease in the steepness during the latter half of the simulation suggests
that even after a many fold increase in the bubble count, the slope may 
reduce only slightly from its
present value and may stabilize at $-8/3$ or thereabouts. 
We have repeated the simulation runs
to assess different realizations and find the above trend consistently evident.
It may be worth mentioning, as an example, that the slope
approached the Kolmogorov value ($-11/3$) after about 10000 bubbles, 
by which time about 42\% of the
volume was occupied by the voids. At the half-way mark and at the end of this 
simulation run the 
occupancy was about 53\% and 54\%, respectively. 
A similar trend was apparent in the spectra
of distributions in Z-slices of the simulated cube. 
The slope here was consistently shallower than
that for the column density distribution (i.e. for a {\it thick} slab), 
by about 1 in the power-law index, as expected.

In the above simulations, the sharp edges of the bubbles are expected to enhance, somewhat
artificially, the power at higher spatial frequencies, and this would make the spectra
shallower than we would find in the absence of sharp edges.
To assess this expectation, we modeled the voids with a smooth decrease in the density as
one moved from the shell boundary to void center. Exactly as anticipated, the spectral 
slopes were consistently steeper (e.g. Figure~\ref{fig:bubble_spectrum}b)
than for those derived for bubbles with sharp-edges 
(e.g. Figure~\ref{fig:bubble_spectrum}a). 
On the average, the power-law slope approached $-3.4$ for the column-density 
distribution and was shallower by one order, i.e. about $-2.4$, for the density in the Z-slices.
For both cases, i.e. the sharp and smooth bubbles, the simulations were repeated
with an additional constraint wherein the resultant distribution of radii was maintained
statistically uniform (as would be implied by a constant SN rate and SNR birth rate). 
The resulting distributions  were
indistinguishable in their appearance from the ones generated without the constraint.
More importantly, the derived spectra (and their slopes) were practically unaffected by 
the additional constraint, implying that the results of these {\it snap-shot} simulations
are not sensitive to the distribution of radii of the bubbles populating the volume. 
In these snap-shot simulations, we have treated SNRs as bubbles "without shells". 
Since the shells surrounding bubbles contribute significant HI column densities, we have
simulated distributions consisting of only shells (with a thickness to radius ratio of 0.1),
and examined the corresponding power spectra. Here, the power-law slope approaches
$-3$ for the total column density distribution, and $-2$ for the density in the Z-slices.

\noindent{\subsection{Time Evolution}}

Following the procedure described in section 3, many simulation runs
were carried out to study the results of a time-evolution approach. 
Due to memory and computing speed
limitations, we restricted the simulation volume to a (128)$^3$ pixel array, and 
we artificially
raised the SN rate by a factor of 100 from the rate appropriate for a (200 pc)$^3$
volume. The primary phase of the runs was chosen long enough so as to have every pixel
of the volume visited by the ejecta. The nominal time scale of this primary phase
was a few million years, and the SN count was several hundred. 
During the ``evolution-only" phase, corresponding to 
a relatively longer duration of time when compared with the typical intervals 
between explosions, the {\it rms} velocity reduced monotonically as shell diffusion 
continued without any further energy input.

Figure~\ref{fig:te_cd_spec} shows the spatial power spectrum of a sample column 
density distribution (such as in Figure~\ref{fig:col_den_sne_image})
obtained from these simulations. 
Although a single power-law is a poor 
approximation to this spectrum, its average (and the best-fit) power-law slope 
interestingly is close to $-11/3$ and does not change in any significant 
manner during the `evolution only' phase. 
Spectra of distributions of column density in
thin slabs (not shown here) reveal a similar non-`single-power-law' nature, 
and they show little variation in the `evolution only' phase. 
However, their average slope is consistently shallower than $-11/3$,
and closer to $-8/3$. Power spectra for several slabs (at different Zs) 
were averaged together to obtain statistically improved estimates.

It is also instructive to examine the velocity distribution and its spectrum. 
In Figure~\ref{fig:te_zv_spec}, we show a sample Z-velocity distribution in the X-Y plane at an arbitrary Z 
and an average of the spatial power spectra of such distributions at various Zs 
(computed separately and then averaged). A single power-law seems adequate to
describe the spatial spectra of 1-d components of the 3-d velocity in our simulations.
The power-law slope is remarkably almost always $-11/3$, or very close to it, 
and may change only slightly through
the `evolution only' phase as the velocity spread (range) 
decreases by orders of magnitude.
While forming these conclusions, we have of course ignored the behavior at the
lower spatial frequency end due to its poor statistical significance.

The parameter and its distribution that can be most closely related to 
observational results is the column density as a function of velocity, as in the
case of, for example, the 21-cm HI line observations. From the simulated distribution
of mass and its 3-d velocity, we extract the spatial distribution 
(in, say, the X-Y plane)
of the column density (i.e. integrated along Z) associated with a given range of the
velocity component along Z. In general, we examine about 30 velocity ranges
covering the estimated spread ($\pm 1\sigma$) of the 1-d velocity, 
and for each of these velocity slices, the transverse column-density distribution 
is Fourier analyzed and the 
power spectra are estimated. Such spectra for the different velocity slices 
are averaged together with natural weighting. 
Figure~\ref{fig:te_vslice_spec} shows a sample
column-density distribution for one such velocity slice 
(across a velocity width of about 1 km/s) 
and a power spectrum obtained by averaging power spectra for different slices.
The spectrum has a slope of $-8/3$ at the higher spatial frequency end. 
It is important to mention
that the same spectral slope also extended to lower spatial frequencies prior to
entering the `evolution only' phase. Such velocity-sliced column-density
distributions appear to show considerable variation in their spectral signatures, 
in general, and more systematically as the `evolution only' phase progresses. 
This is in contrast
to the generally stable signatures evident for the other spectra we discussed above.
Here, almost invariably, the average power-law slope is close to $-8/3$ or shallower. 
The latter is rare during the main-run phase, but is almost always the case during the 
`evolution only' phase. We have investigated this
aspect in greater detail and find that in this phase the spectrum becomes 
progressively shallower, often accompanied by considerable flattening toward 
lower frequencies.  We attempt to understand this in the following way. 
As mentioned earlier, the velocity spread ({\it rms})
decreases monotonically through the `evolution only' phase in the absence 
of any further energy input. As the velocity dispersion decreases and 
mixing continues, the scale over which significant velocity coherence 
survives also decreases progressively. This naturally
leads to significant additional corrugation of the transverse 
distribution on correspondingly smaller spatial scales when viewed 
through a velocity filter that is narrow compared to
the velocity spread. In the present case, the velocity filter widths 
are chosen as a fixed fraction of the {\it rms} velocity, and thus may help 
to reduce the corrugation effects, although the consequences of progressive shortening 
of the coherence scale seem unavoidable.

\noindent{\section{Discussion}}

Before we discuss our results and their implications, it is important to 
mention that although our primary motivation related to the neutral atomic
hydrogen distribution in the ISM, the attempted modeling is equally
relevant for the ionized hydrogen component. In our SN-driven
ISM, it is not far fetched to consider the existence of these two components
as mutually exclusive in their spatial distributions. 
For example, the HI voids correspond to the very
space that is ionized as a consequence of the SN event. HII regions and the thin
HI shells around them (that are in turn surrounded by H$_2$ shells, i.e. molecular 
clouds) represent another example of the
somewhat mutually exclusive existence of the two
components. As long as we are concerned about only the structural features
and spatial distributions, modeling of either of the two components
would have implicit and direct relevance for the other. In more practical terms,
voids/bubbles could be replaced by spheres, and the answers in terms of
spatial spectrum or structure function would be unaltered.

As was seen earlier, the density spectrum from our `snap-shot' simulation approaches a
power-law slope of about -3.4 as the number of supernovae, and the consequent 
bubble-like SNRs, is increased, so long as the bubbles are modeled without 
unnaturally sharp edges. The results of our `time evolution' modeling reveal a spectral
signature of the 1-d velocity distribution that matches remarkably well with that
expected from Kolmogorov turbulence.  
This result appears significant and, in our opinion, is not
a chance coincidence. We remain however aware that though an incorrect
spectral signature might be taken to imply an incorrect scenario,
correct spectral signature does not necessarily imply the (sole) correct scenario. 

 A Kolmogorov spectrum is theorized to
arise when energy cascades from larger scales down to smaller scales.
Implicit in this, we believe, is a certain hierarchical 
structure on a range of spatial scales.
Our results from the snap-shot simulation seem to suggest that
the spectrum actually derived from observations
may primarily be the manifestation 
of structural features that are almost entirely dominated
by the expanding voids/shells of SN/SNRs and/or structures with similar morphology.  
This interesting result may also suggest
that the role any explicit interaction between the ejecta from nearby
events plays is probably not very crucial in defining the statistical distribution.
Though the explicit interaction between nearby events has to occur,
it may contribute significantly mainly at relatively larger spatial-scales,
i.e. as in the formation of hot, connected tunnels (Cox \& Smith, 1974).

The simple approach in our `time evolution' models also seems adequate
to capture the key features of the evolution and the resultant structural
description of the ISM. Each new supernova event is incorporated by introducing 
a {\it seed shell} with its velocity field determined 
based on the mass and the energy input contributed by the SN. 
It is remarkable that  momentum conservation alone
as a governing feature in the subsequent evolution of a given shell 
as it interacts with the surrounding matter seems to yield velocity
structure that has a spatial spectrum similar to that for Kolmogorov turbulence.
This is not surprising, since momentum conservation is considered to be 
the characteristic of the later phases (of SNR evolution) which last the longest.

Although prompted by limitations in computing speeds to use SN rates
that were artificially raised, we have performed elaborate tests to verify that
our basics results, i.e. the spatial spectral signatures, remain valid for a 
wide-range of SN rates. We have also checked and found that the exact shape
of the SN-induced seed shell is not very critical, so long as sharp edges and corners
are not present, the SN statistics are adequately large, and time spans are long enough
to allow enough interaction with the surrounding matter.

We now turn to our specific results. 
Firstly, we find that our simple
simulations provide a more realistic
structural description of the ISM compared to when the distributions are 
synthesized directly from a red spectrum across a spatial frequency range of interest 
(following a suitable power-law).
The present simulations are able to naturally produce 1-d and 2-d features such as
filaments and curved sheets in a manner that is internally consistent with 
the diffuse component and the overall spatial spectrum. 
In fact, such lower dimensional features apparent in our simulated density 
distributions may be contributing to the deviations of the associated
spatial spectrum from a single power-law nature.
Also, any mutual correlations between the velocity and density structures 
is implicit in our `time evolution'
simulations, an aspect that is rather difficult to incorporate by imposing explicit
correlation properties, particularly since it could introduce an undesired bias.

Secondly, it is instructive to note the significant differences between the
spectral descriptions for the column density distributions and for
the distribution of 1-d components of the 3-d velocity. Despite the latter having 
a consistent single power-law with Kolmogorov index, the former clearly deviates
from it at both the low and high spatial frequencies. This may provide important
clues for our understanding of apparent deviations from the Kolmogorov
spectrum that interstellar scintillation observations have indicated
(e.g., Roberts \& Ables 1982), 
prompting serious need to invoke either steeper spectra (i.e., index $\le-4$),
or those with suitably larger `inner scale' (see, for example, 
Goodman \& Narayan 1985; Blandford \& Narayan 1985;
Romani, Narayan \& Blandford 1986).
If our results are to be taken at their face value,
it appears to us that the spectral behavior needed (for column density
across thick or thin screens) to explain the observations
may not be in conflict with the underlying turbulence being of a Kolmogorov
nature. Indeed, if the SN-induced turbulence is to be the dominant process
driving the ISM, there may not be any compelling need for invoking deviations
from a basic Kolmogorov signature. The argument can be turned around to conclude
that the refractive scintillation observations are consistent with, 
and lend support to, 
the possibility that the ISM may be primarily SN-driven.

Thirdly, we draw attention to our results on the distribution
of density across narrow velocity slices. These reveal spatial spectra that
are invariably significantly shallower than $-11/3$, 
and rather close to $-8/3$ or shallower,
even while the spatial structure in velocity remains closely Kolmogorov.
These results can be readily compared with observations, and are found to indeed
be consistent with results derived from the distributions of Galactic
HI emission and opacity observed in narrow frequency (or velocity) channels.
The unavoidable implications of such shallow spectra for the apparent
column density (or opacity) fluctuations on small transverse spatial scales
have been discussed in detail elsewhere (Deshpande 2000). 
For the present discussion we 
focus on the fact that the expected magnitude of such fluctuations
on a given transverse scale has a direct dependence on both the overall 
HI column-density (or opacity) and the details (such as the power-law slope)
of its spatial spectrum. In this particular context, the noticeable
variation in the spectral characteristics associated with the velocity slices
is of great significance. That is, we should not be surprised to find
a considerable variety in the distributions sampled by velocity slices,
and consequently in the
strength of column density or opacity fluctuations while observing
in different directions in the Galaxy. It may be recalled that the 
more significant and systematic modifications of these spectra are seen to occur
as our `evolution only' phase progresses. Had we used the expected SN rate
for the volume simulated (i.e. instead of an elevated rate), 
we would expect the intervals between successive supernovae to be 
long enough to be describable as genuine episodes of `evolution only' phase. 
Given this, it is only reasonable to expect a rich variety of spectral characteristics
to be evident episodically even while observing a given region of the Galaxy.
In practice, for observations made during a narrow epoch range, this would be
the case across the different regions of the Galaxy. Thus, the observed
spatial variation of opacity in the case of even 3C138, often cited  
as evidence for the so-called `tiny-scale structure' (Faison \& Goss 2001, and
references therein), becomes easily understandable
as a simple manifestation of a power-law distribution of opacity across narrow
velocity slices.

Lastly, we will next extend our simple recipe such as to monitor and study
other aspects characterizing the ISM. The effect of Galactic rotation
on the apparent distribution in velocity slices needs to be assessed, particularly
when simulating much bigger volumes than in the present case. More importantly, 
one hopes to keep track of other variables, such as temperature and the resulting ionized
fraction of the ISM.  These extensions will be followed up in a separate paper.

\noindent{\section{Conclusions/Summary}}

In conclusion, we feel that our highly simplified approaches to the modeling of the
ISM have led to several promising and instructive results, as summarized below.

1) Our `snap-shot' simulation of an SN-driven ISM, modeled as a set of 
non-overlapping voids of a range of sizes, appears to be 
successful in capturing the key structural features of the medium, 
including a power-law nature of the spatial spectrum (of slope about -3.4).

2) Another, more realistic, approach following `time evolution' of the 
SN-induced shells also
produces a remarkable match with the structural features of the ISM derived from
available observations. A simple interaction with the surrounding matter,
using only momentum conservation, appears to be adequate to govern the
evolution. 

3) Our simulations show that while the velocity distribution closely follows 
a Kolmogorov spectrum, the column density
distribution shows deviations from this both for high and low spatial frequencies.
This may have important relevance to the properties of the medium as implied 
by observations of refractive scintillation.

4) The spatial spectral characteristics of distributions across velocity slices 
and the significant variation between them suggest that the magnitude of the 
column-density or opacity variation on small
transverse scales can differ considerably when observing different directions 
in the Galaxy.

5) The present simple-minded approaches (i.e. the `time evolution' simulations) 
provide a set of internally consistent 
density and velocity distributions that depict 1-d \& 2-d structure in addition to the 
diffuse component and can be very useful in detailed studies of other properties of the
HI and ionized components of the interstellar medium.

\acknowledgments
We thank Chris Salter, V. Radhakrishnan, Bon-chul Koo and James Ingalls for their 
critical reading of the manuscript and for useful comments.
Our thanks also to an anonymous referee for several critical comments
that have helped us in improving the manuscript. 
We gratefully acknowledge the excellent support from the computer groups at
the Arecibo Observatory and Cal Poly. 
This work was made possible by support from the REU program of NAIC, funded
by the NSF.

\clearpage


\begin{figure}
\leavevmode
\includegraphics[angle=-90,scale=.45]{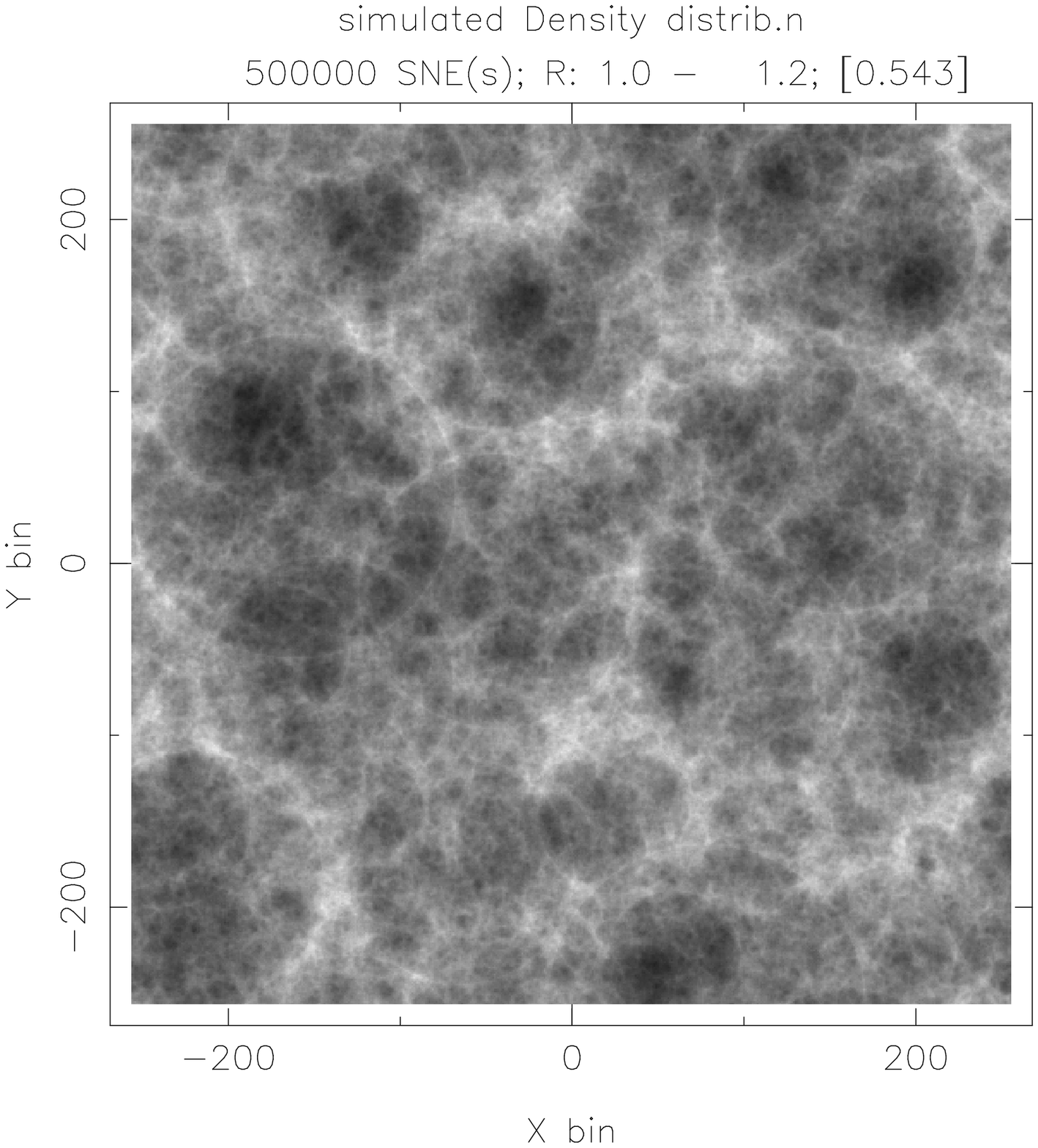}
\hfil
\includegraphics[angle=-90,scale=.45]{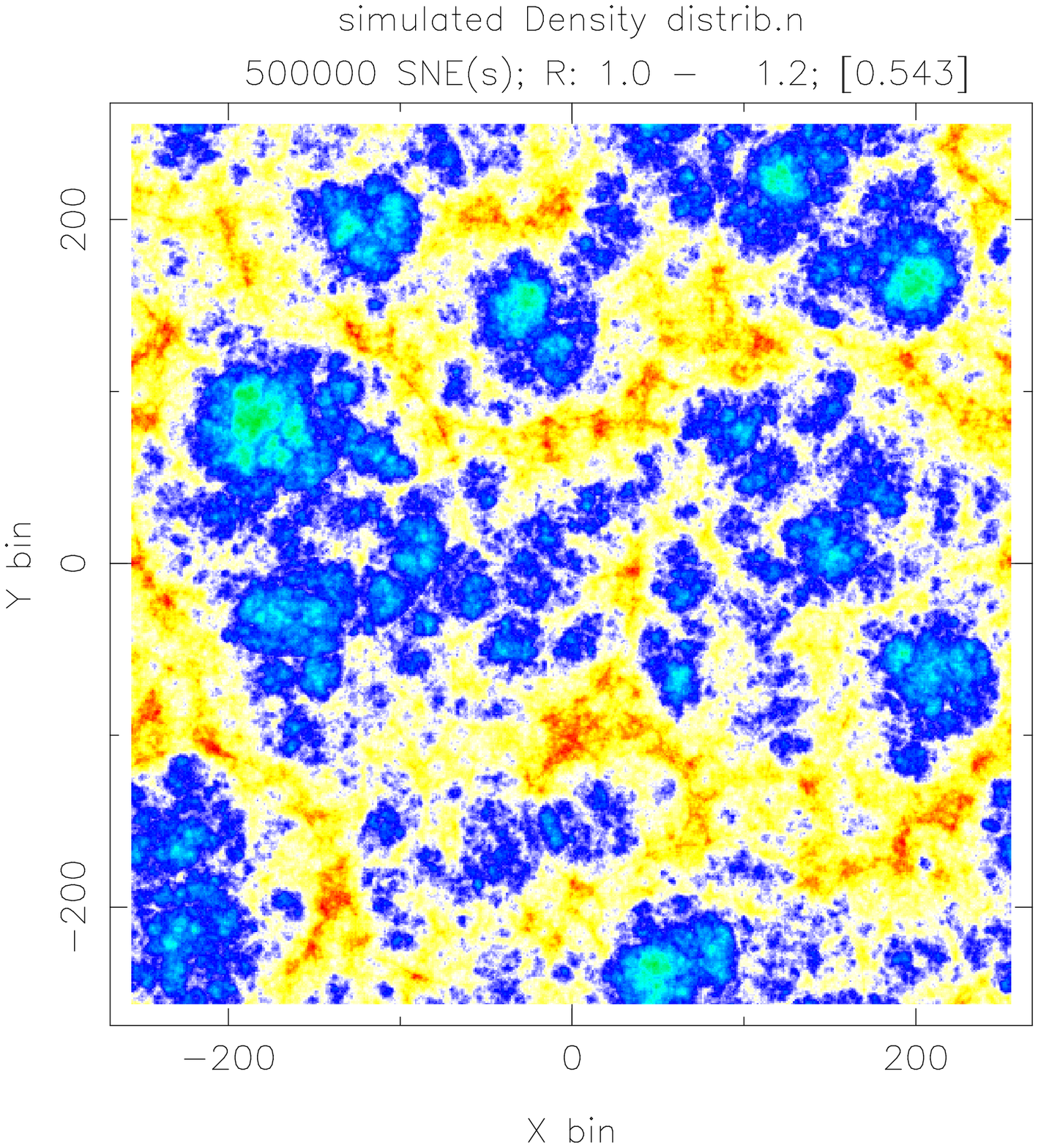}
\caption{The column density distribution resulting from a random distribution of simple 
bubbles/voids with no interaction. This simulation corresponds to 500,000 SN-induced 
bubbles that have randomly chosen sizes and locations within the volume simulated. 
The volume is represented by a (512)$^3$ matrix of cells, and the distribution in x-y,
shown here as an example, is the result of summation along the z-direction. 
The maximum radius allowed for the
bubbles is 0.25 of the side of the cube simulated.
The right panel shows a color plot of the same data, 
where the more diffuse features gain emphasis. 
\label{fig:bubble_image}}
\end{figure}

\clearpage

\begin{figure}
\includegraphics[angle=-90,scale=.8]{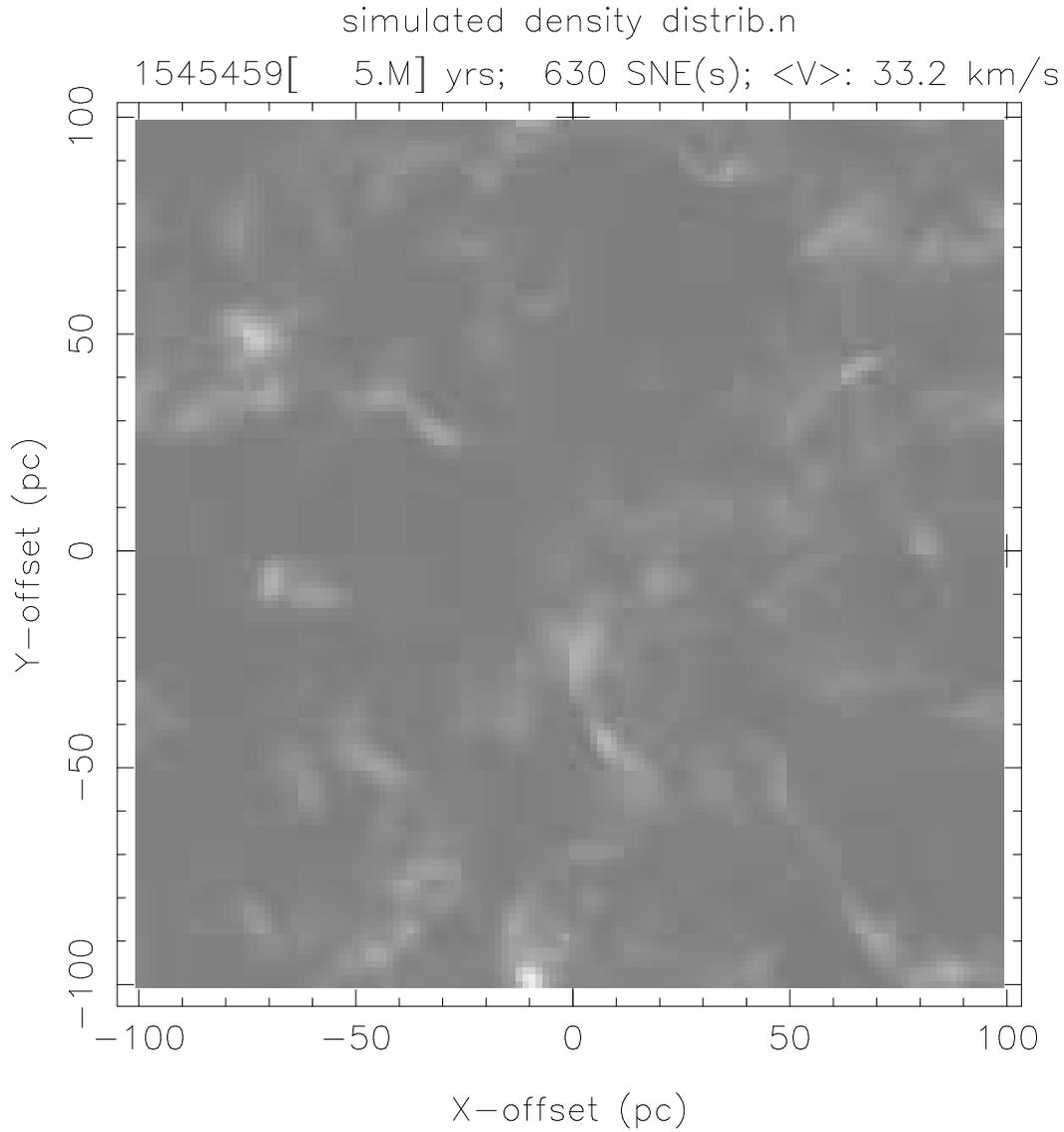}
\caption{Column density distribution resulting from one of our complete 
runs of the ``Time Evolution" simulation. The SN rate was artificially 
raised by a factor of 100.
The first phase time scale was 5 Myr with about 600 supernovae, and 
the extent of the ``evolution only" phase 
was 1.5 Myr. 
\label{fig:col_den_sne_image}}
\end{figure}

\clearpage

\begin{figure}
\includegraphics[angle=-90,scale=.450]{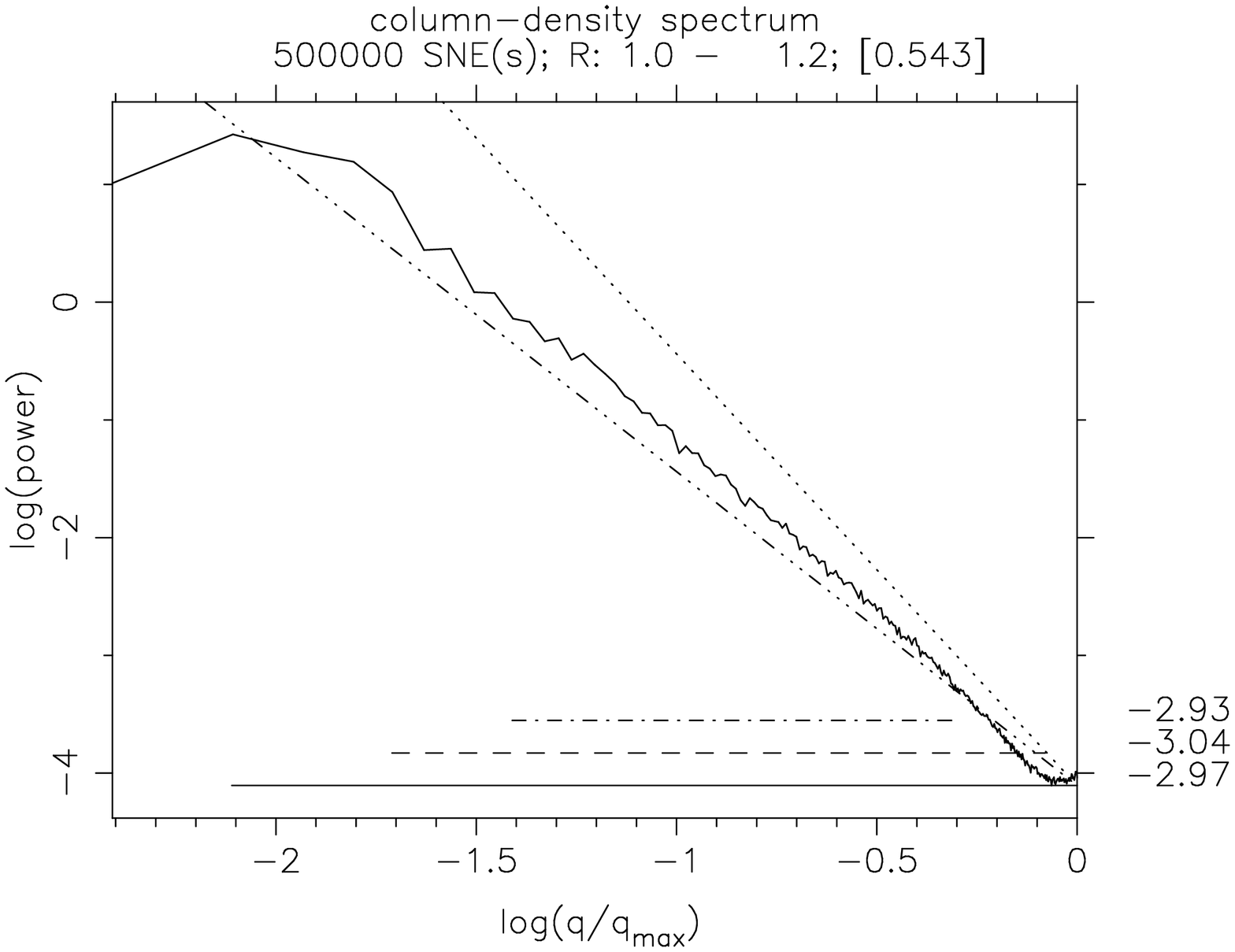}
\vspace{10pt}
\vfil
\includegraphics[angle=-90,scale=.450]{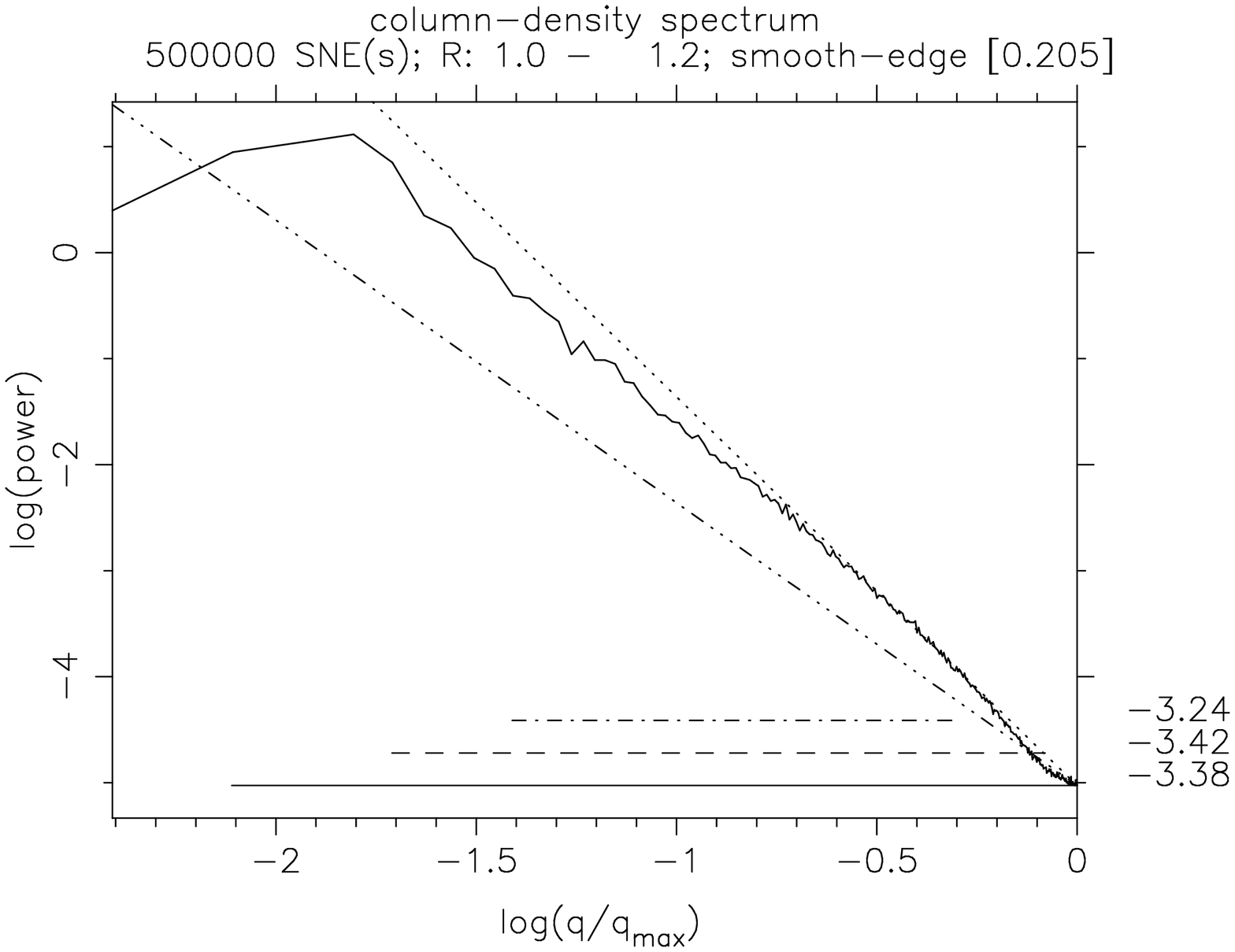}
\caption{The azimuthally-averaged power spectrum of the simulated 
column density distribution shown in 
Figure~\ref{fig:bubble_image}.
The spectral power-law slope in the straight portion of the spectrum 
is about $-3$, i.e. steeper than $-8/3$.
The bottom panel shows a similar result, but from simulations
that use bubbles with smoothed edges.  Here the power-law slope is higher, i.e. about $-3.4$
(see text for discussion). For comparison, two lines with slopes $-8/3$ and $-11/3$ are 
shown in different line styles. The best fit values of slopes of a given spectrum are 
computed for three ranges of the ordinate. These are displayed at the 
bottom right of the panels
and the corresponding ranges are indicated by the horizontal lines at 
the bottom of the panel.
Similarly, we have followed these conventions in plots of 
power spectra elsewhere in this paper.
The value of $q_{max}$, the maximum spatial frequency, is about $(4 pc)^{-1}$ 
in all the spectra shown in this paper. The typical RMS uncertainty in the observed slopes
is about $\pm 0.05$, as estimated from several runs corresponding to a given simulation.
\label{fig:bubble_spectrum}}
\end{figure}

\clearpage

\begin{figure}
\includegraphics[angle=-90,scale=.8]{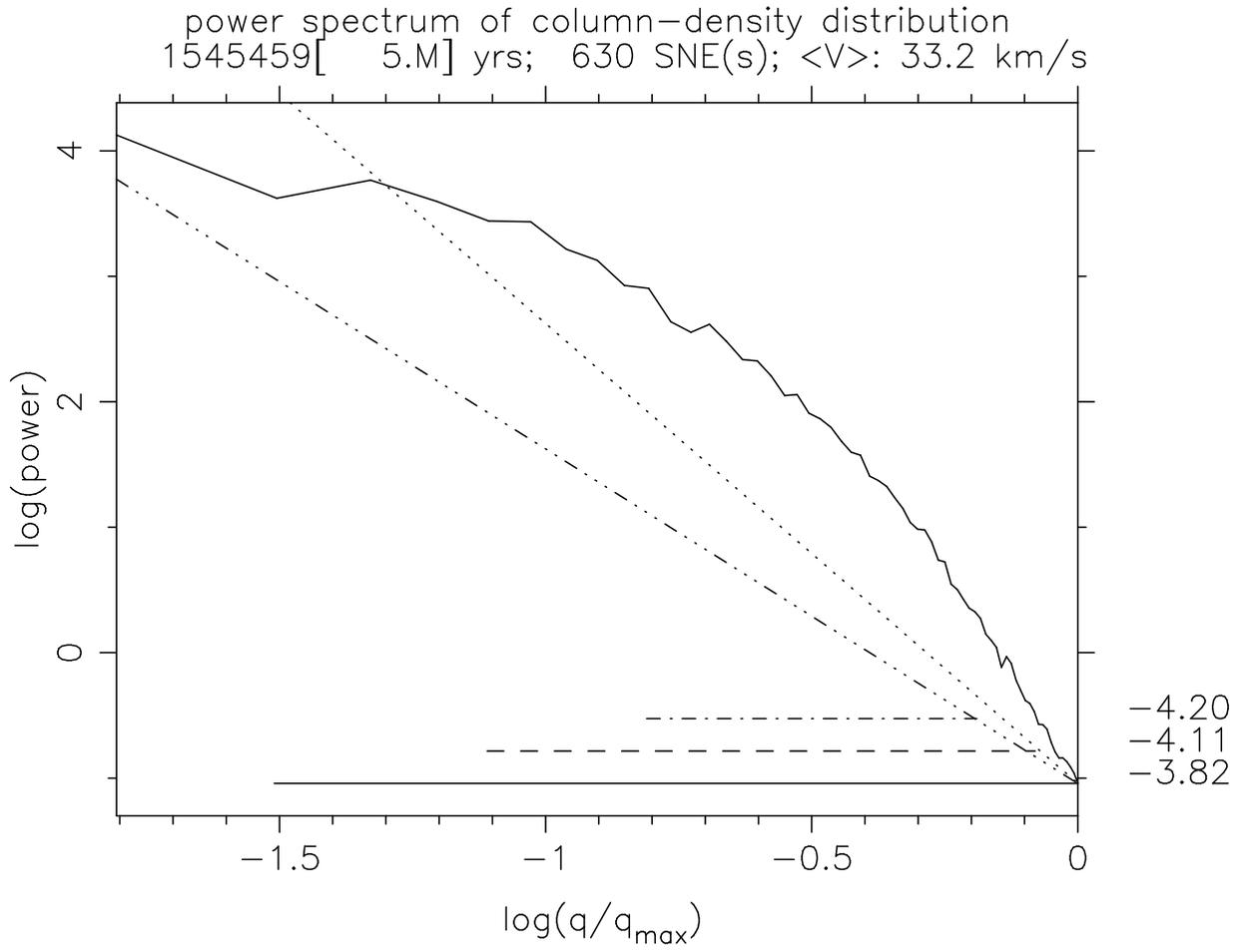}
\caption{Power spectrum of the column density distribution from the `time evolution'
simulation. 
\label{fig:te_cd_spec}}
\end{figure}

\clearpage

\begin{figure}
\includegraphics[angle=-90,scale=.50]{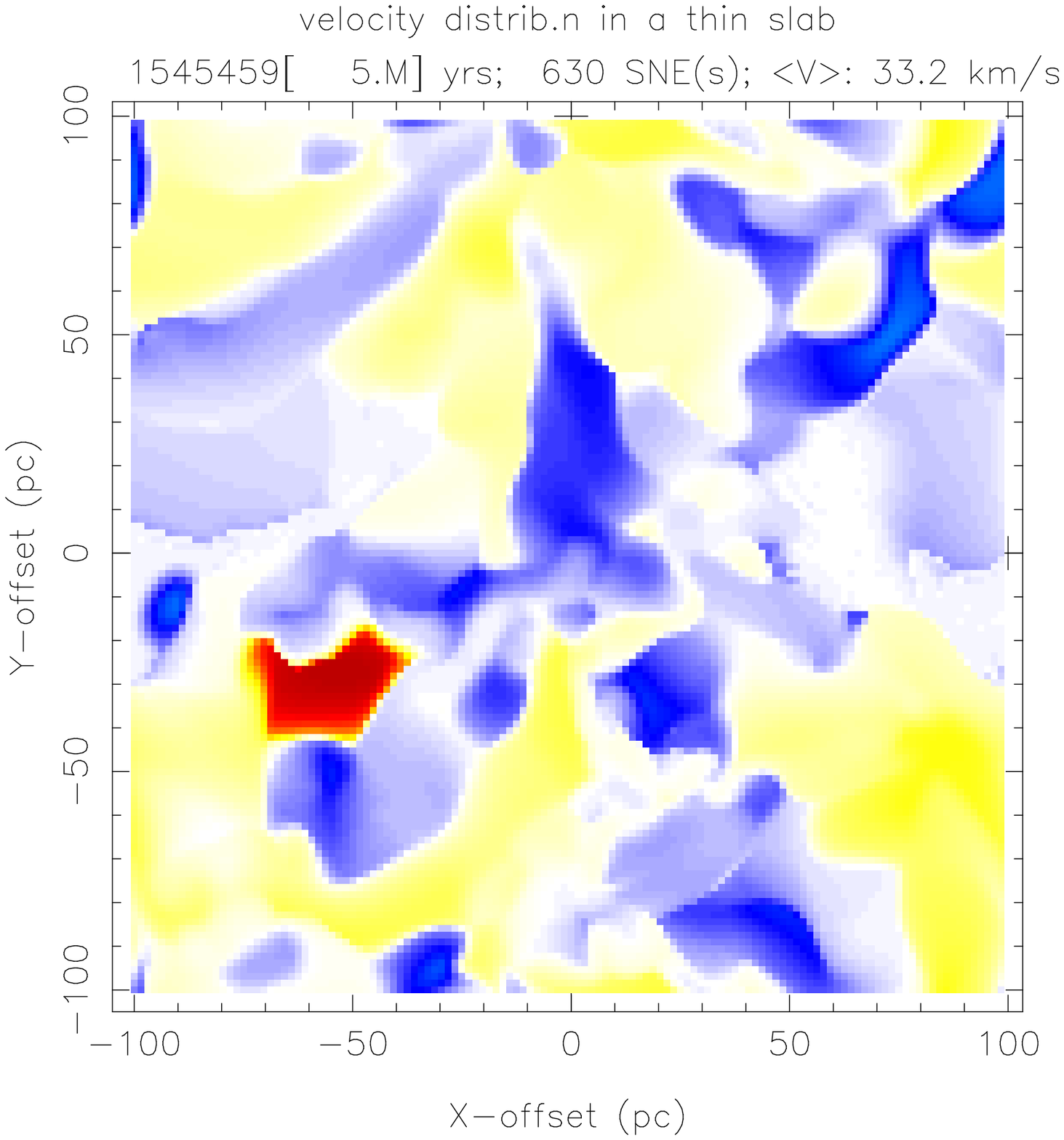}
\vspace{10pt}
\vfil
\includegraphics[angle=-90,scale=.50]{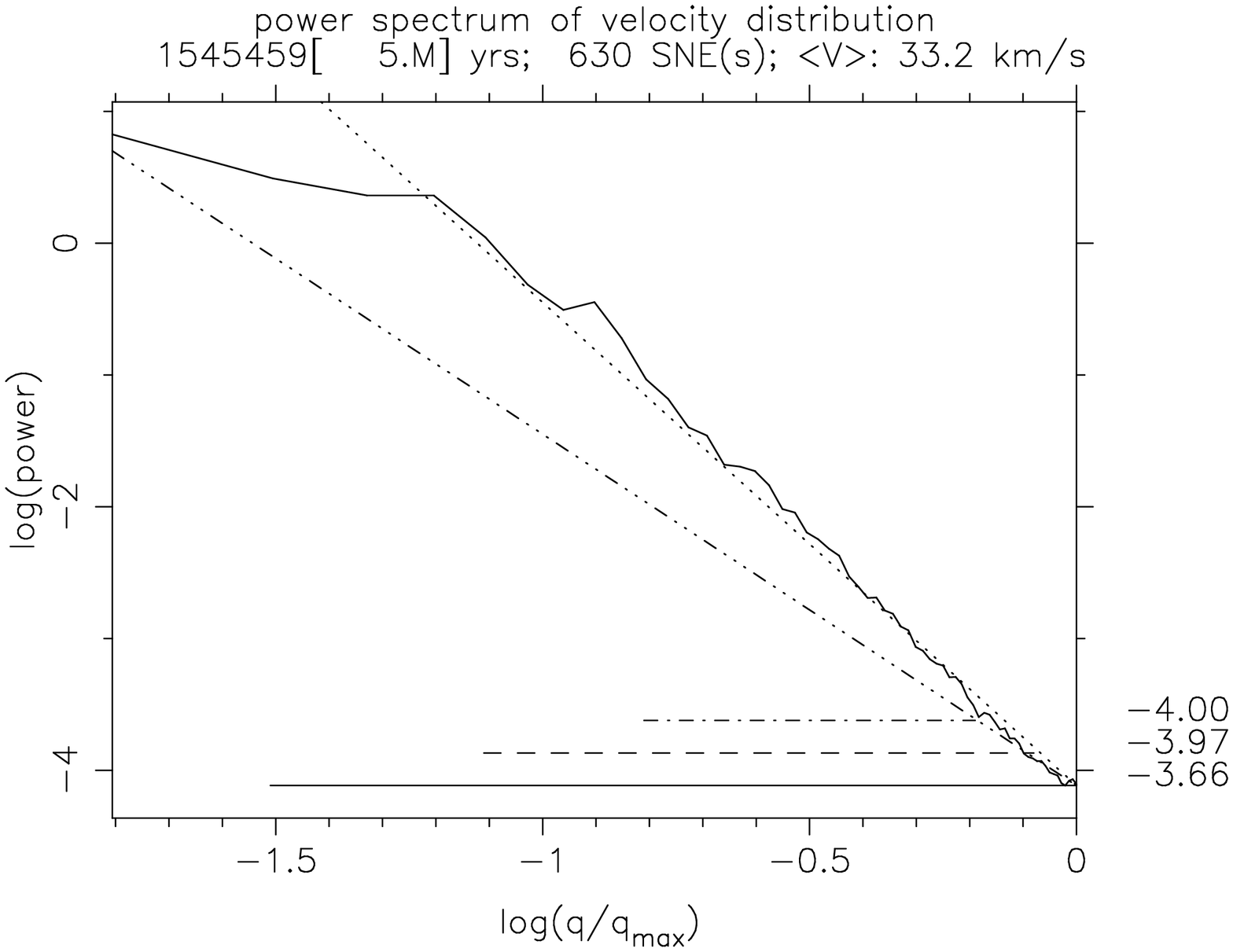}
\caption{A sample `1-d velocity' distribution is shown in the top panel (yellow and red
depict +ve velocities, and the light and dark blue correspond to -ve velocities).
The RMS magnitude of the corresponding 3-d velocity is about 30 km/s.
The spectrum in the bottom panel is obtained by averaging power spectra of several such
distributions.
\label{fig:te_zv_spec}}
\end{figure}

\clearpage

\begin{figure}
\includegraphics[angle=-90,scale=.50]{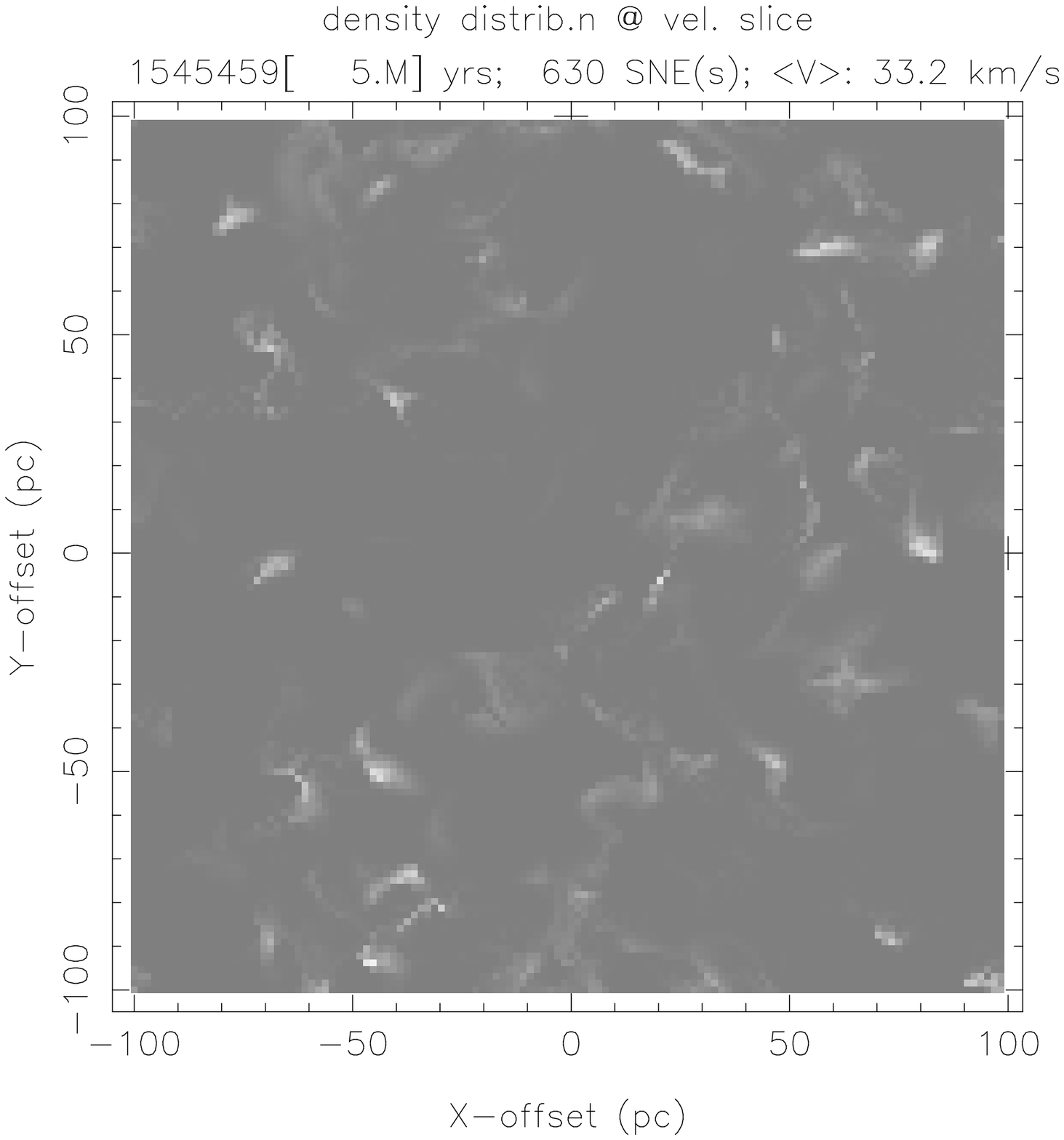}
\vspace{10pt}
\vfil
\includegraphics[angle=-90,scale=.50]{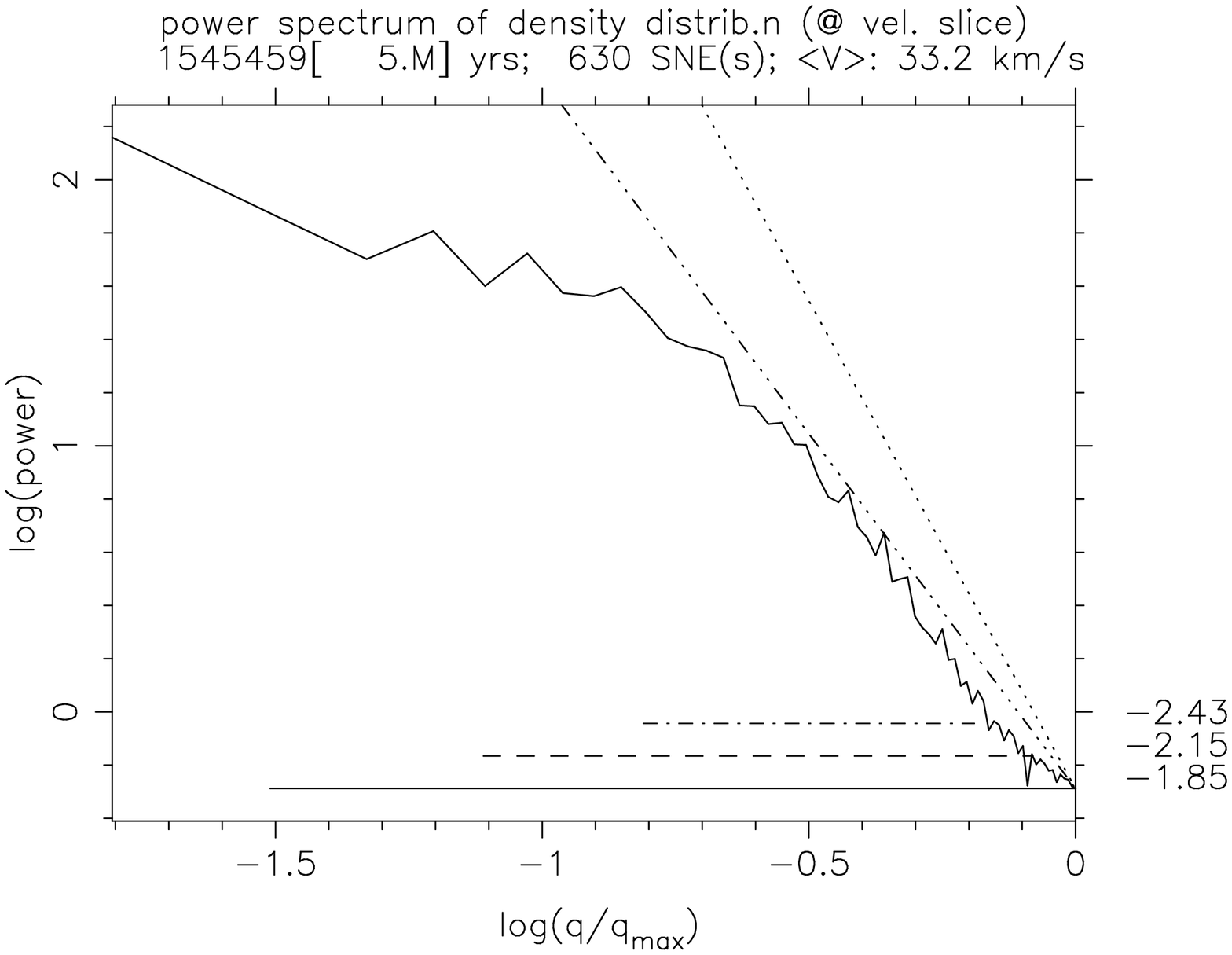}
\caption{A sample distribution of column density associated with a velocity slice, 
with a width of about 1 km/s around zero-velocity, is shown in the top panel.
The spectrum in the bottom panel is obtained by averaging power spectra of distributions as 
would be observable in different velocity channels.
\label{fig:te_vslice_spec}}
\end{figure}


\end{document}